# Detecting a Nuclear Fission Reactor at the Center of the Earth

*R. .S. Raghavan*
*Bell Laboratories, Lucent Technologies, Murray Hill NJ*
*and*
*INFN Laboratori Nazionali del Gran Sasso, Italy*

A natural nuclear fission reactor with a power output of 3-to 10 Terawatt at the center of the earth has been proposed as the energy source of the earth's magnetic field. The proposal can be *directly* tested by a massive liquid scintillation detector that can detect the signature spectrum of antineutrinos from the geo-reactor as well as the *direction* of the antineutrino source. Such detectors are now in operation or under construction in Japan/Europe. However, the clarity of both types of measurements may be limited by background from antineutrinos from surface power reactors. Future U.S. detectors, relatively more remote from power reactors, may be more suitable for achieving unambiguous spectral and directional evidence for a 3TW geo-reactor

The underlying mechanism of the earth's magnetic field and its long-term variability, major topics of geophysical structure, has been debated for decades. A central question is the identity of the energy source needed to produce the field and its variability over periods of $\sim 2 \times 10^5$ years. The traditional model visualizes a core of crystalline iron that grows at a slow rate within an outer core of molten metal. The latent heat of fusion of iron metal and the geodynamical motions in the molten core are pictured as the sources of the energy of the earth's magnetic field and its variability.

A radically different geophysical model of the core structure of the earth envisages an inner core of nickel silicide ($NiSi_x$) rather than iron.[1] In the typically oxygen poor regimes in which $NiSi_x$ forms, the actinides sink to the center to form a kernel concentrated in U and Th. In the early earth 5Gy ago, the $^{235}U/^{238}U$ isotopic ratio was much higher than at present. Thus, conditions exist in the U/Th kernel for a self-sustaining fission reactor. With time, the fast neutron breeding from $^{238}U$ and $^{232}Th$ increases the fission power and sustains it; on the other hand, the build up of fission products and other "poisons" reduces the power until their segregation out of the fission volumes restarts the power rise. Simulations suggest that the model can sustain 3-10 TW of fission power,[2] variable over long time periods, that could be source of a variable earth's magnetic field  The detection of fission specific nuclides could indirectly support the idea of a geo-reactor. Tritium ($^3H$) is produced in a fission reactor. With a low neutron absorption cross section and high mobility, $^3H$ can escape out of the fission volumes and decay to $^3He$ in neutron poor volumes where the $^3He$ can survive. As a stable noble gas, it can percolate eventually to the earth's mantle and beyond. Simulations[2] of this process yield high $^3He/^4He$ ratios, indeed as found in Hawaiian volcanic lavas.[3]

While the NiSi model of the inner core and consequent conclusions of a functional geo-reactor appear self consistent, the evidence is still circumstantial at best and likely to advance only incrementally for the foreseeable future. Because of the implications for geophysical and planetary structure, *direct* evidence for an operating geo-reactor at the center of the earth is highly desirable.  Fortunately, a direct search for a geo-reactor can be carried out in principle, in detection devices planned for other studies.

An operating fission reactor emits signature radiation of *antineutrinos* ( $\bar{\nu}_e$ ) that can be detected at sites remote from the reactor regardless of intervening mass. The detectors can, in principle, also detect the *direction* of the $\bar{\nu}_e$ source. Major detectors are in construction/operation in Europe/Japan for solar neutrino research that are also programmed for $\bar{\nu}_e$ spectroscopy for measuring the global distribution of U and Th in the earth's crust [4] via $\bar{\nu}_e$ emitted in their decay as well as $\bar{\nu}_e$ from *surface* nuclear power reactors for observing neutrino oscillations. It is the signals from the latter that produce the main background against signals of the same type from a geo-reactor. Clear separation of the two signals depends on their relative and absolute rates. In this Note I consider the signal and background problem in such devices from the standpoint of unambiguous detection a 3 –10 TW geo-reactorat the center of the earth.

The science and technology of detecting antineutrinos ($\bar{v}_e$) is well established. Numerous experiments have been carried out using the basic detection reaction: $\bar{v}_e$ + p(proton) → $e^+$(positron) + n (neutron). The visible energy of the positron $e^+$ signal (kinetic + annihilation energy) directly yields the energy of the detected $\bar{v}_e$: $E(\bar{v}_e) = E(e^+) + 0.78$ MeV. The detector is typically a large mass of liquid scintillator—an aromatic organic liquid that emits optical photons proportional to the energy deposited by the ionizing particle. The hydrogen atoms in the liquid serve as the target protons for the above reaction. The special merit of the $\bar{v}_e$ reaction above, is that the $e^+$ signal can be *tagged* as an antineutrino event by a second confirming signal. That signal is provided by the neutron that is thermalized by collision with hydrogen atoms and diffuses in the liquid till, after a delay of several tens of microseconds, it is captured by another proton. The capture produces the second signal. This delayed coincidence tag suppresses the background enormously. The minimum detectable $\bar{v}_e$ flux $\phi(\bar{v}_e)_{min}$ in a well-designed detector with sufficiently low noise from radioactive contamination of the detector (designed e.g., for solar neutrino detection) is then set mainly by the occasional chance delayed coincidence signals.

State of the art detectors are built on the scale of a kiloton of scintillator mass (~ $10^{32}$ protons). The chance coincidence rate in a well designed detector of this type can be limited to several events/year, corresponding to a sensitivity limit of $\phi(\bar{v}_e)_{min}$ ~ $10^4$/cm$^2$ s. The $\bar{v}_e$ flux from a 3-10 TW geo-reactor at any point on the surface is $\phi(\bar{v}_e)_{geo}$ ~ 1-3x$10^5$ /cm$^2$ s. Thus, detecting a geo-reactor by $\bar{v}_e$ spectroscopy is a valid proposition in principle; in practice, the real *non-geo-reactor* $\bar{v}_e$ signals determine the detectability of a geo-reactor.

The spectrum of $\bar{v}_e$ signals from known terrestrial sources is seen in Fig. 1 for liquid scintillation (LS) detectors with a typical mass of a 1 kiloton (with $10^{32}$ protons) located at different sites of the world. The $\bar{v}_e$ signals arise from three main sources: 1) operating commercial power reactors within several 1000km; 2) the distribution of U and Th in earth's crust which differs significantly from a continental site to one surrounded by the ocean; and 3) the postulated geo-reactor at the center of the earth that creates the same signal at any point on the earth's surface. The geo-U/Th spectra for typical geophysical models of the U/Th distribution in the earth's crust and in the mantle have been studied previously.[4]

The $\bar{v}_e$ spectra from the known sources 1) and 2) are shown in Fig 1 in heavier lines while the hypothetical geo-reactor spectra are shown in thinner lines for two different power outputs of 3 and 10 TW. The minimum $\bar{v}_e$ signal energy is 1.02 MeV=$2m_ec^2$. The geo-U/Th signal cuts off at a maximum of ~2.5 MeV ($E(\bar{v}_e)$ =3.3 MeV, from the highest β- energy decay in the U and Th decay chains). Spectra from fission reactors extend beyond, upto ~8 MeV. Signals of the two sources thus appear in separate windows. The reactor spectra and $\bar{v}_e$ fluxes are calculable from the fissile fuel composition, the thermal power (1WTh = 3.11W electrical power) and the distance of the reactor to the detector. A 1 GWTh fission reactor produces 1.786x$10^{20}$ $\bar{v}_e$/s, thus, a flux of 1.31x$10^3$ $\bar{v}_e$/cm$^2$s at a detector 1000 km away. Nominally, a geo-reactor is spectrally nearly *indistinguishable* from power reactors on the surface. (At this stage, we do not exploit spectral differences due to the unknown fuel mixture in a present day breeder geo-reactor (see below)). The geo-reactor offers however, the strong signature of directionality; any surface signal originates from a direction 90° away from the geo-reactor. According to present geophysical models, even the U/Th decay signals in Fig. 1 originate mostly from the earth's crust, nearly at the surface relative to a georeactor.

The signals from a liquid scintillation device, unlike in a Cerenkov detector, are isotropic. In the two step ($\bar{v}_e$,p) reaction above, the $e^+$ signal is followed by the neutron signal displaced in time and location after the n diffusion process. The initial neutron direction vector is kinematically correlated to the neutrino vector. Despite the thermalization and diffusion of the neutron, the $e^+$-n spatial displacement vector still retains the memory of the original neutron direction, thus also the incident neutrino vector. This major result was demonstrated in a practical liquid scintillation detector recently in the CHOOZ experiment[5] which observed the $\bar{v}_e$ spectrum from a 3GW reactor ~1km away. They showed that the data could point to the known direction of the reactor within a cone of half-angle of 18°. The conditions for achieving the result were very favorable. Apart from good energy and position resolution of the signals (achievable routinely in such devices), the reactor signal was strong and clean, with a sample of ~5000 events with a background of only a few percent. Thus in principle, with reasonably good signal/background and signal rates, $\bar{v}_e$ signals from surface power reactors and geo-U.Th could be unscrambled from those of a georeactor in the center of the earth by their orthogonal directions of origin even in a liquid scintillation detector.

Fig.1 surveys the site-related backgrounds against geo-reactor signals in existing and possible future detectors. Panels (a,b) in Fig 1 refer to existing detectors. Borexino, with 300 tons of LS ($0.18 \times 10^{32}$ protons) is under construction in the Gran Sasso underground laboratories in Italy, aimed at solar neutrino studies.[6] It is expected to operate in 2003. A prominent feature of the $\bar{\nu}_e$ spectrum in Borexino is the geo-U/Th signal mainly from the Eurasian continental crust (a small part comes also from the earth's Mantle). The high energy part of the $\bar{\nu}_e$ spectrum arises from European power reactors with a total power of ~450GWTh at average distances of ~800 km mostly *north* from Gran Sasso. Against this background, the geo-reactor signal is comparatively weak. At the lower power of 3TW, it is ~5 times weaker, thus offering little hope for its detection in Borexino, especially with the relatively low signal rates. At 10 TW a $2\sigma$ hint of a geo-reactor signal may be possible in one year's live time. A directionality analysis could only marginally improve matters mainly because of the low signal rate. A future upgrade of the target mass to ~$10^{32}$ protons would help considerably in improving the capability of Borexino for detecting a geo-reactor.

Prospects for the geo-reactor may not be better in the Kamland detector (panel b) in Fig 1), now in operation with a 1 kT LS ($1 \times 10^{32}$ p), in the Kamioka mine in western Japan.[7] The geo-U/Th signal/target proton here is lower (than in Borexino) because Japan is located at the edge of the Asian continent but the adjoining Pacific oceanic crust contributes little. The dominant signal however, comes from Japanese power reactors at an average distance of ~200 km with ~160 GWth total power. Indeed, Kamland is designed to study just these signals to search for $\bar{\nu}_e$ flux disappearance due neutrino oscillation on the 200 km baseline. Against this background, the geo-reactor, even at 10 TW produces a signal only ~15% of the surface reactor signal. Still, the much higher signal rate is an advantage; actual directional analysis of data, available in principle imminently, will clarify to what extent the small georeactor signal is separable from the dominant surface signal. The very possibility of a geo-reactor signal of uncertain magnitude however, possibly creates a caveat on the flux normalization vital for assessing the disappearance effect due to neutrino oscillations. It may thus be necessary to *settle the geo-reactor question even for particle physics objectives* in long baseline reactor experiments such as Kamland.

A map of operating reactors worldwide shows that sites with reactor backgrounds significantly lower than the above cases are likely only in the western U.S. I consider some of the U.S. underground sites that are under discussion for a new National Laboratory for Underground Science (NUSL), such as the Homestake S.D. gold mine and cavities at Carlsbad N.M. (panel c) in Fig.1). (A third site at San Jacinto near the CA coast is unfortunately only ~70 km from powerful reactors. The $\bar{\nu}_e$ spectrum at San Jacinto is thus nearly identical to that at Kamland). The spectra for Homestake and Carlsbad are nearly identical with the favorable feature that the reactor signal /target proton is ~1/3 that at Borexino. A 1 kiloton detector at either U.S. site, would see ~33-110 geo-reactor events/year vs a background of 55-60/y from US reactors. In one year of operation, a $3\sigma$ detection of a 3TW geo-reactor may thus be possible in principle. An advanced state-of-the art 1kiloton (or more) LS detector at one of these proposed NUSL sites thus offers the best initial prospects for directly detecting a geo-reactor since they offer the best combination of signal rates and signal contamination that could facilitate effective operation of the directionality tag.

Ultimately, the best possible site for a geo-reactor search is Hawaii (panel d) in Fig. 1). This option requires however, construction of a new excavated laboratory. In Hawaii, situated entirely on the oceanic crust with very low geo-U/Th. only the small signal from U/Th deep in the Mantle is visible. The remoteness from populated continents on either side reduces the power reactor signals to a comfortably low level. Thus a 1 kiloton detector at an underground location in Hawaii presents the clearest window for a geo-reactor signal even for power outputs less than 3TW. Indeed, with such low background, the possibility exists for attaining a $\bar{\nu}_e$ spectrum ultimately with sufficient signal/noise and statistical precision to derive information on structural details such as the fuel composition in a breeder geo-reactor which bears vitally on its geophysical evolution.

I wish to thank Louis J. Lanzerotti (Bell Labs) and Sandip Pakvasa (U. Hawaii) for bringing recent articles on the geo-reactor proposal to my notice.

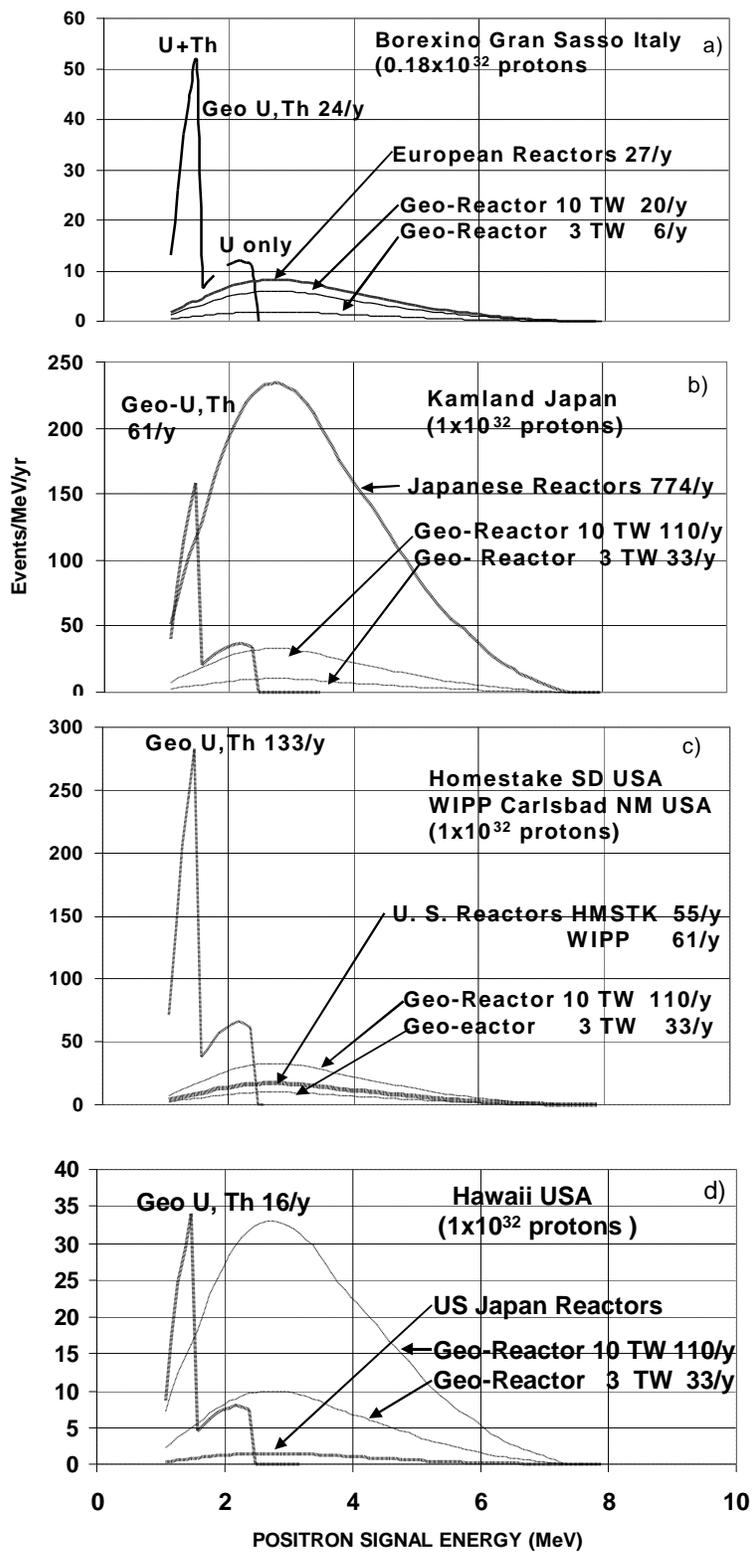

Fig. 1 Antineutrino spectra observable in detectors at several sites in Europe, Japan and the U.S.